\documentclass[epj]{svjour}
\usepackage{graphics}
\begin{document}
\title{Flow effects on jet profiles and multiplicities}
\author{N\'estor Armesto\thanks{\emph{Present address:} Departamento de
F\'{\i}sica
de Part\'{\i}culas and Instituto Galego de Altas Enerx\'{\i}as, Facultade de
F\'{\i}sica, Campus Sur,
Universidade de Santiago de Compostela, 15782 Santiago de Compostela,
Spain.}
}                     
\institute{Department of Physics, CERN, Theory Division,
CH-1211 Gen\`eve 23, Switzerland}
\date{Received: date / Revised version: date}
%
\abstract{
We study the effects of low-$p_T$ collective flow on radiative
energy loss from high-$p_T$ partons traversing the QCD medium created in
high-energy nucleus-nucleus collisions. We illustrate
this idea through three examples. Due to longitudinal flow,
jet profiles at the LHC present marked
asymmetries in the $\eta\times \phi$-plane, and
widths in $\eta$
and $\phi$ of particle distributions
associated with a high-$p_T$ trigger at RHIC become different.
Finally, transverse flow implies an increase of
high-$p_T$ $v_2$ at RHIC.
} 
\maketitle
\section{Motivation and formalism}
\label{intro}
                                                                                
Low transverse momentum inclusive spectra and
azimuthal correlations measured in Au+Au collisions at the Relativistic Heavy
Ion Collider
(RHIC)
indicate that different
hadron species emerge from a common medium which has built up
a strong collective velocity field
\cite{Ackermann:2000tr,Adcox:2002ms,Adler:2003kt,Kolb:2003dz}.
These measurements are broadly consistent
with calculations based on ideal hydrodynamics, whose success is
regarded as stro\-ng evidence \cite{Gyulassy:2004zy}
that the medium produced in nucleus-nucleus collisions
has a very small mean free path, shows a very rapid
thermalization at a time less than 1 fm/c after initial impact,
and behaves like an almost ideal fluid with vanishing viscosity. This suggests
strong position-momentum correlations in the medium.
In this contribution we explore possible effects of
this collective flow on high-$p_T$ observables
\cite{Armesto:2004pt,Armesto:2004vz}.

\begin{figure}
\begin{center}
\resizebox{0.5\textwidth}{!}{%
  \includegraphics{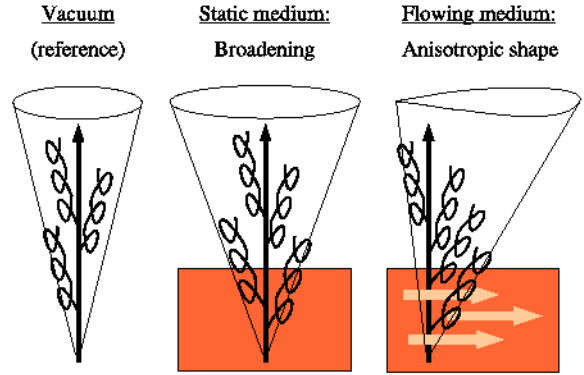}
}
\end{center}
\caption{Sketch of the radiation from a fast parton traversing the vacuum
(left), a static medium (center) and a flowing medium (right).}
\label{fig1}       
\end{figure}
At collider energies, the production of high-$p_T$ hadrons and
jets provides a novel independent characterization of the
produced medium. This is so since
the gluon radiation off parent partons is sensitive to the
interaction between the partonic projectile and the
medium, see the reviews in
\cite{Kovner:2003zj,Gyulassy:2003mc,Salgado:2003qc,Armesto:2004sb,urshere}.
The radiative energy loss of a fast parton traversing a QCD medium is
determined by momentum
exchanges perpendicular to the trajectory of the parton. Thus, if the hard
parton (jet) is produced in a frame not co-moving with
the collective flow, momentum exchanges become anisotropic and an
additional contribution to energy loss comes from flow, see Fig. \ref{fig1}.

At a given energy density $\epsilon$, the dynamic behaviour of
the medium is fully specified by its equation of state (EOS)
$p = p(\epsilon,T,\mu_B)$ which enters the energy momentum tensor
\begin{equation}
  T^{\mu\nu}(x) = \left(\epsilon + p \right)\, u^{\mu}\, u^{\nu}\,
                  - p\, g^{\mu \nu}\, .
  \label{eq1}
\end{equation}
Here, $u^{\mu}=\gamma(1,\vec{\beta})$ is the flow velocity field.

On the other hand, quenched high-$p_T$ hadroproduction
is sensitive to the transport coefficient $\hat{q}$, which is
proportional to the density of scattering centres and characterizes
the squared average momentum transfer from the medium to the hard
parton per unit path length. This transport coefficient is related
to $\epsilon$ \cite{Baier:2002tc},
$  \hat{q} \left[{\rm GeV}^2/{\rm fm} \right]\,
  = \, c \, \epsilon^{3/4} \left[({\rm GeV}/{\rm fm}^3)^{3/4} \right]$.
Here, $c$ is a proportionality constant of order unity
\cite{Baier:2002tc,Eskola:2004cr}.

In order to take into account the effect of the anisotropy in momentum
exchanges due to the presence of flow on radiative energy loss, we modify the
Yukawa-like scattering potential usually employed to model the medium
\cite{Gyulassy:1993hr},
\begin{equation}
\vert a({\bf q}) \vert^2
  = \frac{\mu^2}{\pi \left[ ({\bf q} - {\bf q}_0)^2 + \mu^2 \right]^2}\, .
  \label{eq3}
\end{equation}
The Debye screening mass $\mu$ is usually taken to be proportional to the energy
density $\epsilon$, while the directed momentum component ${\bf q}_0$ is
assumed to come from the additional contribution to $T^{ii}$ is given by
$\Delta p =
(\epsilon + p) u^i\, u^i (= 4\, p\, \gamma^2\beta^2$
for an ideal EOS $\epsilon = 3\, p)$. This implies,
for rapidity differences $\eta =
0.5, 1.0, 1.5$ between the frame co-moving with the hard parton and the frame
co-moving with the collective flow, $\Delta p/p\simeq 1, 5, 18$. So this
naive estimation shows that such directed component may be $\vert {\bf
q}_0\vert \sim \mu$.
Let us indicate that we do not address in this contribution
the dynamical dilution of the medium. It can be taken into
account by a redefinition of the (time-dependent) transport coefficient as
shown in \cite{Baier:1998yf,Gyulassy:2000gk,Salgado:2002cd}.

We have calculated the medium-induced radiation of gluons with energy
$\omega$ and transverse momentum ${\bf k}$, emitted from a highly
energetic parton that propagates over a finite path length $L$ in a
medium of density $n_0$ with collective
motion. To first order in opacity, we
find \cite{Wiedemann:2000za,Gyulassy:2000er,Salgado:2003gb}
\begin{eqnarray}
 \omega \frac{dI^{\rm med}}{d\omega\, d{\bf k}} &=& \frac{\alpha_s}{(2\pi)^2}
 \frac{4\, C_R\, n_0}{\omega} \,
 \int d{\bf q}\, \vert a({\bf q})\vert^2\,
 \frac{ {\bf k}\cdot {\bf q}}{{\bf k}^2}
 \nonumber \\
 && \times
 \frac{-L \frac{({\bf k} + {\bf q})^2}{2\omega} +
       \sin\left(   L \frac{({\bf k} + {\bf q})^2}{2\omega}\right)}
     {  \left[({\bf k} + {\bf q})^2 / 2\omega\right]^2} \, .
 \label{eq4}
\end{eqnarray}
In the absence of a medium, the parton fragments according to the vacuum
distribution $I^{\rm tot} = I^{\rm vac}$. The radiation spectrum (\ref{eq4})
characterizes the medium modification of this distribution
$  \omega\frac{dI^{\rm tot}}{d\omega\, d{\bf k}}
  = \omega\frac{dI^{\rm vac}}{d\omega\, d{\bf k}}
  + \omega\frac{dI^{\rm med}}{d\omega\, d{\bf k}}$ in the
so-called single hard scattering
approximation \cite{Wiedemann:2000za,Gyulassy:2000er,Salgado:2003gb}.
Physically equivalent results are obtained in the multiple soft
scattering approximation \cite{Baier:1996sk}, as explained
in \cite{Salgado:2003gb}.
From $\omega\frac{dI^{\rm tot}}{d\omega\, d{\bf k}}$, we calculate
distortions of jet energy and jet multiplicity
distributions \cite{Salgado:2003rv}.

\section{Exercises}
\label{sec2}
To illustrate the effects of flow on high-$p_T$ observables and the
information about the medium that can be extracted from such measurements, we
perform three exercises: We examine jet shapes at the Large Hadron
Collider (LHC)
\cite{Armesto:2004pt,Armesto:2004vz} in Section \ref{lhc}, widths of
particle distributions at RHIC \cite{Armesto:2004pt} in Section \ref{rhic1}
and,
in Section \ref{rhic2}, high-$p_T$
elliptic flow at RHIC
\cite{Armesto:2004vz}.
                                                                                
\subsection{LHC: jet shapes}
\label{lhc}

From $\omega\frac{dI^{\rm tot}}{d\omega\, d{\bf k}}$, we calculate
distortions of jet energy and jet multiplicity
distributions \cite{Salgado:2003rv}.
Information about $I^{\rm vac}$ is obtained
from the energy fraction of the jet contained in a subcone of radius
$R = \sqrt{\eta^2 + \phi^2}$,
$  \rho_{\rm vac}(R)$.
For this jet shape, we use the parametrization \cite{Abbott:1997fc}
of the Fermilab $D0$ Collaboration for jet energies in the range
$\approx 50 < E_t < 150$ GeV and opening cones $0.1 < R < 1.0$, modified as
explained in \cite{Armesto:2004pt,Armesto:2004vz}.
We then calculate from
Eq. (\ref{eq4}) the modification \cite{Salgado:2003rv} of
$\rho_{\rm vac}(R)$ caused by the energy density and collective flow
of the medium. To do so, we transform the gluon emission
angle ${\rm arcsin}\left(k/\omega\right)$ in (\ref{eq4}) to jet
coordinates $\eta$, $\phi$,
$  k\, dk\, d\alpha = \omega^2\, \frac{\cos\phi}{\cosh^3\eta}\,
  d\eta\, d\phi\, $
where $\alpha$ denotes the angle between the transverse gluon
momentum ${\bf k}$ and the collective flow component ${\bf q}_0$.
\begin{figure}
\begin{center}
\resizebox{0.5\textwidth}{!}{%
  \includegraphics{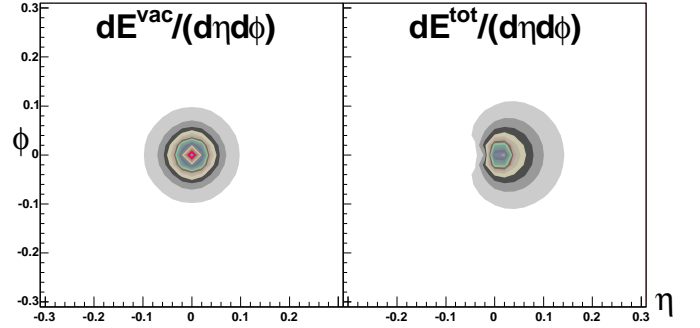}
}
\end{center}
\caption{Calculated
distortion of the jet energy distribution 
in the $\eta \times \phi$-plane for a 100 GeV jet. The right
hand-side is for an
average medium-induced radiated energy of 23 GeV and equal
contributions from density and flow effects, $\mu = q_0$.}
\label{fig2}       
\end{figure}

In Fig. \ref{fig2}, we show the medium-modified jet shape for
a jet of total energy $E_T = 100$ GeV. To test the sensitivity
of this energy distribution to collective flow, we have chosen
a rather small directed flow component, $q_0 = \mu$. The
parameters in (\ref{eq4})
were adjusted \cite{Armesto:2004pt,Armesto:2004vz} such that
an average energy $\Delta E_T = \int d\omega \omega
\frac{dI^{\rm med}}{d\omega} = 23$ GeV is redistributed by
medium-induced gluon radiation.
This is a conservative estimate for Pb+Pb collisions at the
LHC, and implies a shift of the calorimetric jet center
of 0.04 rapidity units \cite{Armesto:2004vz}.
Both the broadening in the jet structure due to medium-induced Brownian
motion of the partonic jet
fragments in a dense medium \cite{Salgado:2003gb} and a marked rotational
asymmetry in the
$\eta \times \phi$-plane characteristic of the presence
of a collective flow field, can be observed. While the jet energy
contours shown in
Fig. \ref{fig2} may not be easy to observe, the
difference in widths in the different directions in the $\eta \times
\phi$-plane should be clearly visible, see
\cite{Armesto:2004pt,Armesto:2004vz}.
                                                                                
\subsection{RHIC: widths of particle distributions}
\label{rhic1}
\begin{figure}
\begin{center}
\vskip -0.3cm
\resizebox{0.4\textwidth}{!}{%
  \includegraphics{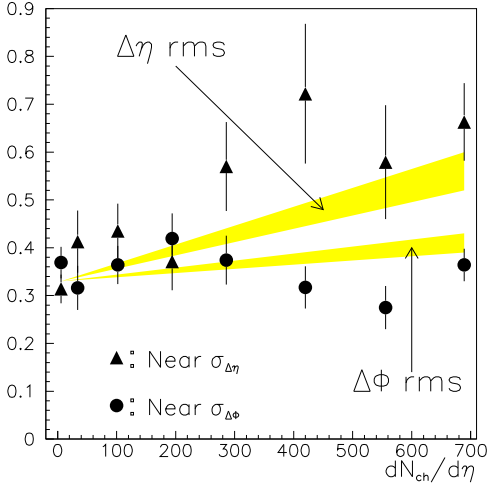}
}
\end{center}
\caption{The width in azimuth and rapidity of the near-side
distribution of charged hadrons associated to high-$p_T$
trigger particles of transverse momentum
$4\, {\rm GeV} <  p_T < 6\, {\rm GeV}$ in Au+Au collisions
at $\sqrt{s_{NN}} = 200$ GeV. Black points are preliminary data
from the STAR collaboration \cite{FuqiangWang}.
The band represents our
calculation for longitudinal flow fields in the range
$2< q_0/\mu < 4$, see text for further details.}
\label{fig3}       
\end{figure}

The calculation of medium-induced gluon radiation is
most reliable for calorimetric measurements, but it also provides a
framework for the discussion of medium-modified multiplicity
distributions \cite{Armesto:2004pt,Armesto:2004vz,Salgado:2003rv}. At RHIC
calorimetric measurements are not currently performed,
and the present discussion
is on particle distributions associated to high-$p_T$ triggers
\cite{FuqiangWang,Wang:2004kf,danhere}. We have compared our calculation to
data
of the STAR Collaboration \cite{FuqiangWang,Wang:2004kf},
which measured the widths of the $\eta$- and
$\phi$-distributions of produced hadrons associated to trigger particles
of transverse momentum $4\, {\rm GeV} <  p_T < 6\, {\rm GeV}$.
As a function of centrality of the collision, the $\phi$-distribution
does not change within errors, while the $\eta$-distribution shows
a significant broadening, see Fig. \ref{fig3}.
Although these data are still preliminary, they allow us to illustrate
the strategy of determining collective flow effects from jet asymmetries.
To this end, we have first used the width of the jet-like correlation
in p+p collisions to characterize the vacuum contribution.
The energy of the parent parton was fixed to 10 GeV. We
have chosen a rather small in-medium path length $L = 2$ fm
to account for the fact that high-$p_T$ trigger particles tend
to correspond to parent partons produced near the surface.
We then calculated the asymmetry of the broadening in
$\Delta \eta$ and $\Delta \phi$ by varying the average momentum
transfer between $\mu  = 0.7$ and $\mu = 1.4$ GeV, and the
size of the collective flow component between $q_0/\mu = 2$
and $q_0/\mu = 4$. The results thus obtained for central Au+Au
collisions were extrapolated to peripheral ones by a straight line
and are represented by the band in Fig. \ref{fig3}.
Numerical uncertainties in applying calculations of parton energy
loss to transverse hadron momenta $p_T < 10$ GeV are significant
and have been discussed repeatedly \cite{Salgado:2003gb}.
However, the origin of the
angular broadening of jet-like particle correlations is essentially
kinematic, being determined by the ratio between the momentum
transfer from the medium and the energy of the escaping particle;
hence, the result in Fig. \ref{fig3} should not depend strongly
on the details of our calculation.
We observe that the ratio $q_0/\mu = 4$, corresponding to a boost of the
energy-momentum
tensor (\ref{eq1}) by approximately one unit in rapidity $\Delta \eta$,
can account
for the tendency in the preliminary STAR data of Fig. \ref{fig3}.
It is consistent with a space-time picture of Au+Au collisions at RHIC
in which the co-moving frames of the hard parent partons of trigger particles
and of the medium are boosted by one rapidity unit.

\subsection{RHIC: elliptic flow}
\label{rhic2}

In general, a hard parton will suffer less energy loss if it
propagates on a trajectory parallel to the flow field. Thus, for
the same medium-induced suppression, the azimuthal asymmetry at high
transverse momentum becomes larger when the contribution
of the collective flow field is increased. To estimate
the size of this effect, we consider a simple two-dimensional model,
see \cite{Armesto:2004vz} for details.
The hard parton is produced at an arbitrary position $(x_0, y_0)$ in the
transverse plane according to the nuclear overlap. It propagates
in its longitudinally co-moving rest frame in the transverse direction
$\vec{n} = \left( \cos\varphi, \sin\varphi\right)$,
along the trajectory
$  {\bf r_0}(\xi) = \left(x_0 + \xi \cos\varphi, y_0 + \xi \sin\varphi
                    \right)$.
For simplicity, we assume that the longitudinally co-moving rest
frame of this hard parton is the longitudinal rest frame of the
medium. Then, there is only a transverse but not a longitudinal
flow component. In this exercise we will work in the multiple soft scattering
approximation \cite{Salgado:2003gb,Baier:1996sk}.
For the BDMPS transport coefficient which includes
collective flow effects, we make the ansatz
\begin{eqnarray}
  \hat{q}(\xi) = q_{nf} +
                 q_{f}
                 \vert  u_T({\bf r}_0(\xi))\cdot {\bf n}_T \vert^2\, .
  \label{5.9}
\end{eqnarray}
Here $q_f$ and $q_{nf}$ stand
for the flow and non-flow components to $\hat{q}$, the two-dimensional
vector ${\bf n}_T$ is orthogonal to the parton trajectory
and projects out the corresponding transverse component
of the collective flow field $u_T({\bf r}_0(\xi))$.
Defining $q_{nf}$ as the
time-averaged transport coefficient of the dynamically
equivalent static scenario \cite{Baier:1998yf,Gyulassy:2000gk,Salgado:2002cd},
the ansatz (\ref{5.9})
can account for one of the main effects of longitudinal expansion, namely
the time-dependent decrease of the density of
scattering centers. In the presence of
collective flow, there is an additional momentum transfer orthogonal
to the parton trajectory and hence parallel to
$\vec{n}_T  = \left( - \sin\varphi, \cos\varphi\right)$. Since
the transport coefficient denotes the squared average
momentum transfer per unit path length, this contribution
enters quadratically, $\vert u_T({\bf r}_0(\xi))\cdot {\bf n}_T \vert^2$.

For an exploratory model study, we use a
blast-wave parameterization of the hadronic freeze-out stage
of the collision \cite{Retiere:2003kf}, see \cite{Armesto:2004vz} for details.
With this input, we calculate the characteristic gluon energy
and average transverse momentum squared for a parton trajectory
in a medium characterized by its density distribution
and its collective flow field. With the
ansatz (\ref{5.9}) for the BDMPS transport coefficient, we find
\begin{eqnarray}
  \omega_c({\bf r}_0,\varphi) &=& \int_0^\infty d\xi\, \xi\, \hat{q}(\xi)\,
  \Omega({\bf r}(\xi), \xi)\, ,
  \label{5.14} \\
  \left(\hat{q} L\right)({\bf r}_0,\varphi)
   &=& \int_0^\infty d\xi\, \hat{q}(\xi)\,
  \Omega({\bf r}(\xi), \xi)\, .
  \label{5.15}
\end{eqnarray}
$\omega_c({\bf r}_0,\varphi)$
depends linearly on $q_{nf}$ and
on the relative flow strength $q_f/q_{nf}$.
It increases with increasing flow for
parton trajectories which are not parallel to the flow field and
shows distortions \cite{Armesto:2004vz} which, as $\Delta E \approx \alpha_s
\omega_c$ \cite{Baier:2002tc}, provide a first
indication
of the extent to which parton energy loss depends
on a transverse flow field and affects the azimuthal distribution
of inclusive hadron spectra.

\begin{figure}
\begin{center}
\vskip -1.2cm
\resizebox{0.5\textwidth}{!}{%
  \includegraphics{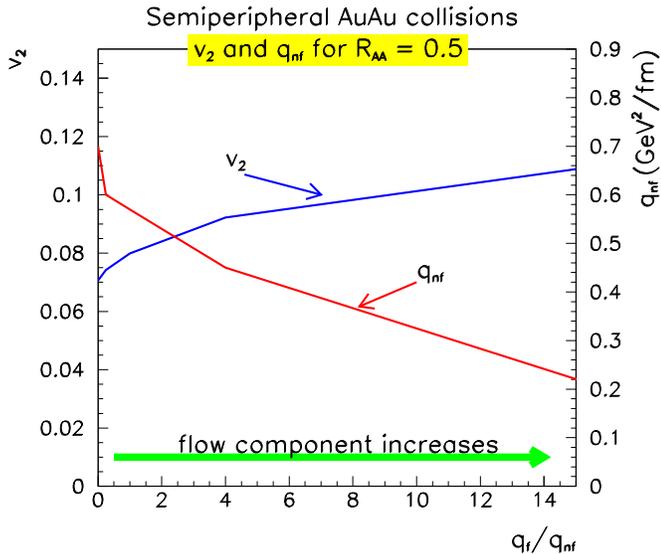}
}
\end{center}
\caption{The dependence of elliptic flow $v_2$ and the non-flow
component of the BDMPS transport coefficient ${q}_{nf}$
on the relative flow strength ${q}_f/{q}_{nf}$, for the
case of a nuclear modification factor $R_{AA} = 0.5$ in
semi-peripheral Au+Au collisions for a
fixed transverse momentum $p_T = 7$ GeV.}
\label{fig12}       
\end{figure}

To estimate the effects of transverse flow, we get
from (\ref{5.14}) and (\ref{5.15}) the relative suppression
of hadronic spectra due to medium-induced parton energy loss
$  N(x_0,y_0,\varphi,p_T)$ $=$
  $\frac{d\sigma^{med}}{dp_T} \big/
  \frac{d\sigma^{vac}}{dp_T}$.
For details of the evaluation, see \cite{Armesto:2004vz}.
The results, see
Fig. \ref{fig12},
illustrate two qualitative effects of transverse flow: First,
low-$p_T$ elliptic flow induces an additional contribution to
high-$p_T$ azimuthal asymmetry. This effect may reduce significantly
the discrepancy of models of parton energy
loss \cite{Drees:2003zh,Dainese:2004te}
in accounting for high-$p_T$ $v_2$. Second, the presence of
collective flow diminishes strongly the local energy density
$\epsilon \propto {q}^{4/3}_{nf}$ of the medium required
for a nuclear modification factor $R_{AA}$ of fixed size.
                                                                                
\section{Summary}
\label{summ}

We have performed an exploratory study of the
effects of collective flow on medium-induced radiative energy loss.
Jets at LHC \cite{Accardi:2003gp} may show a clear
$\eta-\phi$-asym\-me\-try.
At RHIC flow can produce
asymmetries in associated particle
production for different directions, and an
sizable increase of elliptic flow.
The determination of
densities from jet quenching studies becomes more involved, as flow may mimic
energy density.
Theoretical uncertainties \cite{urshere} exist: finite energy
corrections, hadronization (which requires pp and pA data),$\dots$, but
even
a negative
result would provide information about the space-time evolution of the system
(e.g. of the coupling between hard production and flow). Truly quantitative
answers require the computation of this effect
within a full hydrodynamical
simulation \cite{Hirano:2002sc}.

\vskip 0.1cm

\noindent {\bf Acknowledgments:} The work presented here was made
\cite{Armesto:2004pt,Armesto:2004vz} in
collaboration with Carlos A. Salgado and Urs Achim Wiedemann.
I thank R. Baier, P. Jacobs, D.
Magestro,
A. Morsch,
J. Schukraft and F. Wang for useful discussions, and the organizers
for such a nice meeting.
%
%

\end{document}